\documentclass[12pt]{article}

\usepackage{lscape}
\usepackage{multirow}
\usepackage{amssymb,amsmath}
\usepackage{amstext}
\usepackage{amscd}
\usepackage{graphics}
\usepackage{epsfig}

\def\wt{\widetilde}
\newcommand{\Rho}{{\mbox{\sf P}}}

\def\ts{\textstyle}
\def\w{{\hspace{.1mm}{\rm w}\hspace{.1mm}}}

\def\be{\begin{equation}}
\def\ee{\end{equation}}
\def\beq{\begin{equation}}
\def\eeq{\end{equation}}
\def\bea{\begin{eqnarray}}
\def\eea{\end{eqnarray}} 

\def\eqn#1{(\ref{#1})}

\def\nn{\nonumber}

\def\sideremark#1{\ifvmode\leavevmode\fi\vadjust{\vbox to0pt{\vss
 \hbox to 0pt{\hskip\hsize\hskip1em
 \vbox{\hsize3cm\tiny\raggedright\pretolerance10000
  \noindent #1\hfill}\hss}\vbox to8pt{\vfil}\vss}}}

\begin{document}
\thispagestyle{empty}

\vspace{.8cm}
\setcounter{footnote}{0}
\begin{center}
\vspace{-25mm}
{\Large
 {\bf Weyl Invariance and the Origins of Mass}\\[5mm]

 {\sc \small
     A.R.~Gover$^{\mathfrak G}$, A.~Shaukat$^{\mathfrak S}$,  and A.~Waldron$^{\mathfrak W}$\\[4mm]

            {\em\small ${}^{\mathfrak G}\!$
            Department of Mathematics\\ 
           The University of Auckland,
            Auckland, New Zealand\\
            {\tt gover@math.auckland.ac.nz}\\[1mm]  
            ${}^{\mathfrak S}\!$
            Physics Department\\  
            University of California,
            Davis CA 95616, USA\\
            {\tt ashaukat@ucdavis.edu}
            \\[1mm] 
            ${}^{\mathfrak W}\!$
            Department of Mathematics\\ 
            University of California,
            Davis CA 95616, USA\\[-2mm]
            {\tt wally@math.ucdavis.edu}
            }}

 }

\bigskip

{\sc Abstract}\\[-4mm]
\end{center}

{\small
\begin{quote}

By a uniform and simple Weyl invariant coupling of scale and matter
fields, we construct theories that unify massless,
massive, and partially massless excitations.  Masses are related to
tractor Weyl weights, and Breitenlohner-Freedman stability bounds in
anti de Sitter amount to reality of these weights.  The method relies
on tractor calculus -- mathematical machinery allowing Weyl invariance
to be kept manifest at all stages. 
The equivalence between tractor and higher spin systems with
arbitrary spins and masses is also considered.
\end{quote}
}


\section{Introduction}

Local symmetries are the guiding principle to construct many physical theories.   Electromagnetism and other Yang--Mills theories stand witness to the success of the gauge principle.  Typically Yang-Mills theories do {\it not} respect local scale, or what is often called Weyl, invariance. Notable exceptions are the conformal gravity and  supergravity theories of~\cite{Kaku:1977pa} which are based on Yang--Mills theories of the conformal, or superconformal groups\footnote{Indeed there is an intimate relation between the ``gauging of spacetime algebras'' methodology of~\cite{Kaku:1977pa,Kaku:1978nz,Townsend:1979ki,PvN}
and the tractor calculus techniques that are presented here.}.
Also, non-Weyl invariant theories can be turned into  Weyl invariant ones by using Weyl compensators~\cite{Zumino,Deser0}.  
Our aim is to use Weyl invariance as a principle to construct theories.

Actually, lightlike  theories in constant curvature backgrounds have been constructed using 
Weyl invariance in~\cite{Deser:1983tm}.  There lightlike propagation was achieved by
imposing the sufficient (but not necessary) condition of rigid conformal invariance.
Those theories were constructed in two steps: (i) The models were
coupled Weyl invariantly to an arbitrary, conformally flat, metric
when both the metric and the physical fields transformed.  (ii) The
metric was held constant while only the physical fields transformed,
which turned the original Weyl invariance into rigid conformal
invariance.  In this Letter we extend this method to incorporate both
massless and massive theories (lightlike or not) in curved backgrounds
and as a consequence uncover a relationship between mass and Weyl
invariance.

This relationship relies on an elegant description of Weyl invariance
in terms of mathematical objects called tractors which are used for
studying conformal geometry~\cite{East,Bast,Rice,Slov}. Tractors are
to Weyl invariance what tensors are to diffeomorphism invariance.
They are classified by their tensor type, and ``tractor weights''
under Weyl transformations.  Analogously to the
situation in diffeomorphism invariant theories, there are relatively
few tractors and tractor operators; this allows us to efficiently
engineer Weyl invariant theories.

Our methods\footnote{A more detailed and complementary description of
  our work may be found in~\cite{GoSW}.} yield massive, massless and
partially massless~\cite{Deser:1983tm,Higuchi:1986py,Deser:2001pe}
theories in a unified way.  As is well known, mass
emerges at the cost of broken Weyl invariance which seemingly
contradicts our proposal. Here, as a single principle, we posit that
the breaking of Weyl invariance is via a Weyl invariant interaction of
the matter field with the scale field; the interaction is mediated by
a Weyl covariant operator applied to the scale yielding the so-called
scale tractor below (see also \cite{almost,GDN}). 
Associated to Weyl invariance is a certain
parabolic subgroup of $SO(d,2)$; as we shall see below the scale
tractor is precisely the object which, pointwise, reduces the
parabolic to $SO(d-1,1)$.  
To understand the breaking of Weyl
invariance, let us review and reformulate an old idea called Weyl
compensators~\cite{Zumino}.

\section{\label{dilaton} Weyl Compensators and the Scale Tractor}
Theories with dimensionful couplings are actually gauge fixed versions of Weyl invariant ones. For example, the Einstein-Hilbert action
$
S_{\rm EH}(g_{\mu\nu})=-\frac1{2\kappa^2}\int \! d^dx\,\sqrt{-g} \, R
$,
with coupling $\kappa$ comes from a
Weyl invariant action
\be
S(g_{\mu\nu},\sigma)=-\frac{1}{2}\int d^dx\, \sqrt{-g}\, \sigma^{-d}\Big(\,(d-2)(d-1)[\nabla_\mu\sigma]^2+R \, \sigma^2\Big)\, ,\label{conf_inv_act}
\ee
upon identifying the dimension one field $\sigma = \kappa^{\frac{2}{d-2}}$.
Since
$
S(g_{\mu\nu},\sigma) = \kappa^2 S_{\rm EH}(\sigma^{-2}g_{\mu\nu})$, 
it is Weyl invariant when both $g_{\mu\nu}$ and  $\sigma$ transform as
\be
\label{Wtrans}
g_{\mu\nu} \mapsto \Omega^2 g_{\mu\nu}\, ,\qquad 
\sigma \mapsto \Omega \sigma  \,   .
\ee
The {\it r\^ole} of the Weyl compensator/dilaton $\sigma$ is to guarantee Weyl invariance.

An alternative way to represent~\eqn{conf_inv_act} in a manifestly Weyl invariant way is to use $\sigma$ to build a new $(d+2)$-dimensional vector~$I^M$
\be
I^M = \ \begin{pmatrix}
\sigma \\[1mm] \nabla^m \sigma \\[1mm] -\frac 1 d [\Delta +\frac{R}{2(d-1)}]\sigma\, \end{pmatrix} \, ,
\ee
which, under Weyl transformations~\eqn{Wtrans}, transforms as 
\be
I^M \mapsto   U^M{}_N I^N \, , \label{tractorvector}
\ee 
where the $SO(d,2)$ matrix $U$ is given by (when $e_{\mu}{}^{m}\mapsto \Omega e_{\mu}{}^{m}$)
\be U=
\begin{pmatrix}
\Omega&0&\;0\;\\[2mm]
\Upsilon^m&\delta^m_n&\;0\;\\[2mm]
-\frac12\Omega^{-1}\,\Upsilon_r\Upsilon^r&-\Omega^{-1}\Upsilon_n&\Omega^{-1}
\end{pmatrix}\, , \qquad \Upsilon_\mu = \Omega^{-1}\, \partial_\mu \Omega\, .
\label{Um}
\ee Parabolic $SO(d,2)$ transformations of this special form will be
called ``tractor gauge transformations''.  Objects that transform
like~\eqn{tractorvector} are called weight zero tractors and $I^M$,
being a special tractor, is called the scale tractor. On the other
hand it is evident that $I$ yields a reduction of the parabolic in
$SO(d,2)$ to $SO(d-1,1)$; this motivates its use as the tool for
mediating scale interactions. 

Recasting the action in tractor language, it takes the manifestly Weyl invariant form:
\be
S(g_{\mu\nu},\sigma)=\frac{d(d-1)}2\int \frac{\sqrt{-g}}{\sigma^d}\,  I^M \eta_{MN} I^N \label{Isq} ,
\ee
where $\eta_{MN}$ is the block off-diagonal, $SO(d,2)$-invariant, tractor metric 
\be
\eta_{MN}=\begin{pmatrix}0&0&1\\0&\eta_{mn}&0\\1&0&0\end{pmatrix}\,
.\label{eta} \ee 
The scale tractor is doubly special:  It introduces the Weyl compensator or a ``scale'' $\sigma$ in the theory and controls the breaking of Weyl invariance. Holding the scale constant both picks a metric and yields the 
dimensionful coupling~$\kappa$ which allows dimensionful masses to be calibrated to dimensionless Weyl weights.    

{\it Bona fide} Weyl invariant (massless) theories are now described
in an interesting way: At special Weyl weights the scale tractor
automatically decouples  from the matter fields. 

 This
mechanism underlies the existence of locally scale invariant theories
in special dimensions and/or with very particular couplings --- the
classic example being four-dimensional Maxwell theory.

In the following, we construct massless and  massive theories using a single framework.  
Tractors are the basic mathematical tools for this unification;  a short introduction is in order (see~\cite{G,GP,CG} for a
 detailed account).

\section{\label{sec:level1}Tractor Calculus}
Under (local) Weyl transformations, the metric transforms by 
\be
g_{\mu\nu} \mapsto \Omega^{2}(x)\,  g_{\mu\nu}\, .\label{Metric} 
\ee 
This is the local symmetry of conformal geometry whose building blocks are the vielbein, Levi-Civita connection, and the $Rho$-tensor, a clever combination of which produce the  ``tractor connection" and its curvature\footnote{Here $W_{\mu\nu}{}^{mn}$ is the Weyl tensor, $C_{\mu\nu}{}^{m}$ is the Cotton tensor which is in turn the covariant curl of the $Rho$ or Schouten tensor $\Rho_{\mu\nu}\equiv\frac1{d-2}\Big(R_{\mu\nu}-\frac12\frac1{d-1}g_{\mu\nu}R\Big)$.
 }
 \be {\cal D}_\mu=
\begin{pmatrix} \partial_{\mu} &-e_{\mu n}&0\\[2mm]
\Rho_{\mu}{}^m&\nabla_{\mu}&e_{\mu}{}^m\\[3mm]
0&-\Rho_{\mu n}&\partial_{\mu} 
\end{pmatrix}\, ,
\quad
{\cal F}_{\mu\nu}=
\begin{pmatrix}
0&0&0\\[2mm]
C_{\mu\nu}{}^m&W_{\mu\nu}{}^m{}_n&0\\[2mm]
0&-C_{\mu\nu n}&0
\end{pmatrix}\, .
\ee
This connection transforms  under  Yang-Mills transformations
$ {\cal D}_\mu\mapsto U\ {\cal D}_{\mu} U^{-1}$
with $U$  given by~\eqn{Um}. It is fundamental to tractor calculus.

We define a weight $w$ tractor vector by the transformation law 
\be T^M\equiv
\begin{pmatrix}
T^+\\T^m\\T^- 
\end{pmatrix}\mapsto\Omega^w U^M{}_N T^N \, . \nn\\\label{tractortrans}
\ee 
To raise and lower tractor indices $M,N\ldots$, we can use the weight zero, symmetric, rank two tractor metric in~\eqn{eta}
as it is parallel with respect to~${\cal D}_\mu$.
Weight $w$ tractors can be mapped covariantly to $w-1$ ones by  the   $D$-operator
\be
D^M\equiv\begin{pmatrix}(d+2w-2)w\\[2mm](d+2w-2){\cal
  D}^m\\[2mm]-({\cal D}_\nu{\cal D}^\nu + w \Rho)\end{pmatrix}. \label{D}
\ee 
Actually, acting on a weight one scalar field $\sigma$, the $D$-operator  produces  the scale tractor 
$
I^M = \frac{1}{d}D^M\sigma
$.
On conformally Einstein manifolds, $I^M$ is parallel with respect to ${\cal D}_{\mu}$ and $[D^M, I^N] = 0$.  

The $D$-operator does not obey a Leibniz rule but there is an integration by parts formula when the Weyl weight of the integrand is zero.  For example, if $V^M$ is a (compactly supported) weight $w$ tractor vector and $\varphi$ is a weight $1-d-w$ scalar then 
\be \int \sqrt{-g} \, V_M D^M \varphi = \int \sqrt{-g}\,
\varphi D^M V_M\, .\label{parts} 
\ee 
Another useful invariant tractor operator is 
$
X_M\equiv\begin{pmatrix} 1\; \ 0 \; \  0\end{pmatrix}
$,
which can be used to access the top component/slot of a tractor vector.
Both the aforementioned operators are null,
$D^M D_M = 0=X^MX_M$.
Lastly, we define a further tractor operator built from these, called
the Double $D$-operator~$D^{MN}$,~by
\be 2 D^{MN}= [X^M, D^N] + (d+2w)\eta^{MN} \,
. \label{identity} \ee 
We now have enough tractor technology to start
building physical theories.

\section{\label{sec:level1}Scalars}
Massive scalar field theory in curved backgrounds can be described using tractors.  
Suppose $\varphi$ is a weight $w$ scalar field transforming as
$ \varphi
\mapsto \Omega^w \varphi$
and let us  search for a Weyl covariant, tractor equation of motion for ~$\varphi$. A massive theory is expected to involve the scale tractor  but  further tractor operators are needed to build a scalar equation of motion. 
The most promising is  the   $D$-operator from which we construct the equation of motion
 \be I_M D^M \varphi = 0\, .\label{sceom} 
\ee
In a Weyl  frame where $\sigma$ is constant, the equation of motion becomes
\be
-\sigma\Big(\Delta+\frac{2\Rho}{d} w(w+d-1)\Big)\varphi=0\,
.\label{scomp} \ee
At an arbitrary weight $w$, and spaces with constant~$\Rho$, equation~\eqn{sceom} describes a massive field propagating in a curved background
$
(\Delta-m^2)\varphi=0
$,
 with mass-Weyl weight relation given by
\be
m^2 = \frac{2\Rho}{d}\Big[\Big(\frac{d-1}2\Big)^2-\Big(w+\frac{d-1}2\Big)^2\Big]\, .\label{massweight}
\ee
Since $\Rho$ is negative and constant in anti de Sitter spaces, reality of  the Weyl weight~$w$ implies the Breitenlohner-Freedman bound~\cite{Breitenlohner:1982jf,Mezincescu:1984ev} \be m^2\geq \frac{\Rho}{2d}\,
(d-1)^2\, , 
\ee 
for stable scalar propagation in that case. This also predicts a bound in any space with constant negative~$\Rho$.

At the special weight $w=1-d/2$, the scale $\sigma$ completely  decouples from~\eqn{sceom} and we find a  conformally improved scalar field
$(\Delta-\frac R4\, \frac{d-2}{d-1}\, )\varphi=0$ which is expressed tractorially as
\be D^M \varphi =
0\, .\label{spsc} \ee
The scale tractor $I^M$ decouples because at $w=1-d/2$,  the top and middle slots of the   $D$-operator vanish while the bottom slot is the conformally invariant wave operator.   The same conclusion follows from  the Weyl invariant action 
\be
S[g_{\mu\nu},\sigma,\varphi]=\frac12 \int \sqrt{-g} \, \sigma^{1-d-2w}
\varphi I_M D^M \varphi =
S[\Omega^2g_{\mu\nu},\Omega\sigma,\Omega^w\varphi]\, ,
\ee
whose variation produces~\eqn{sceom}.  At $w=1-d/2$, the scale actually disappears 
because $(\sigma d)^{-1} (D^{M}\sigma) D_{M}$ then equals the conformally improved wave operator. We now construct a single tractor vector theory describing
the spin one Maxwell, Proca and conformally invariant/gauge variant  systems of~\cite{Deser:1983tm}.

\section{\label{Vectors} Vectors}
To write a vector theory of a potential $V_{\mu}$ in terms of tractors, we  introduce scalar fields $(V^+,V^-)$ arranged as a  weight $w$ tractor vector~multiplet \be V^M=\begin{pmatrix}V^+\\ V^m
\\ V^-\end{pmatrix}\, .
\ee
As we have more fields than needed, we impose the Weyl covariant constraint
\be
D^M V_M = 0\, .\label{DV}
\ee
The constraint eliminates $V^-$ as an independent field\footnote{This is true at generic values of $w$. Consult~\cite{GoSW} for details of distinguished values.} leaving the unwanted field $V^+$.  We turn to gauge invariance to convert $V^+$ into a 
St\"uckelberg field
\be
\delta V^M = D^M \xi\, ,\label{eee}
\ee
which, for the independent components, yields
\bea \delta V^+ &=& (d+2w)(w+1)\, \xi\, ,\nn\\[2mm] \delta V_\mu \,
&=& (d+2w)\nabla_\mu \xi\, .\label{maxgauge} 
\eea 
It is clear that for $w$ arbitrary, $V^+$ is an auxiliary field in a St\"uckelberg description  of the massive Proca system while at $w=-1$, it can be consistently set to zero allowing us to recover Maxwell system.

Next, we introduce the gauge  invariant  ``tractor Maxwell curvature''
\be {\cal F}^{MN}=D^M V^N - D^N V^M\, .\ee
To build a vector field equation for $V^{M}$, we must contract ${\cal F}^{MN}$ with another tractor vector; inevitably to describe mass,
this must be the scale tractor
\be I_M {\cal F}^{MN}\equiv G^N
= 0.\label{maxeom} 
\ee 
Picking a Weyl  frame where $\sigma$ is constant and  the background metric 
is Einstein, eliminating the  higher derivatives that appear in equations of motion~$G^{M}$ by  differentiating and taking linear combinations, we find a massive Proca equation
\be
-\sigma \Big(\nabla_n F^{nm} +\frac{2\Rho}{d}(w+1)(d+w-2)\wt V^m\Big)\equiv {\cal G}^m=0\, . \label{Proca}
\ee
Here $
\wt V_\mu = V_\mu -\frac 1 {w+1} \nabla_\mu V^+$
is St\"uckelberg gauge invariant and its coefficient gives  the mass-Weyl weight relationship and Breitenlohner--Freedman bound

\be
m^2=\frac{2\Rho}{d}\Big[\Big(\frac{d-3}{2}\Big)^2-\Big(w+\frac{d-1}{2}\Big)^2\Big]\ \geq\ \frac{2\Rho}{d}\, \Big(\frac{d-3}{2}\Big)^2\, . \label{maxmass}
\ee

Gauging away the St\"uckelberg field and taking divergence of~\eqn{Proca}, we obtain 
the standard on-shell description of massive spin one excitations
\be
\Big\{\Delta +{\textstyle\frac{ {\ts 2}\Rho}{{\ts  d}}}\, [w(w+d-1) -1]\Big\} V_\mu=0= \nabla^\mu V_\mu = 0 \, ,
\label{spin_one_onshell}
\ee
which are neatly expressed tractorially as
$ I\cdot D\ V^M=0=
I\cdot V =
 X\cdot V$.

Let us analyze special values of $w$:  When $w=-1$, as anticipated above, equation~\eqn{Proca} become Maxwell's equations $\nabla_m F^{mn}=0$.
Just as for scalars, at $w=1-d/2$ the scale $\sigma$ decouples and we find the conformally invariant but gauge variant theory of~\cite{Deser:1983tm}
\be
\Delta A_\mu - \frac 4 d \, \nabla^\nu \nabla_\mu A_\nu +\frac{d-4}{d}\Big(2\, \Rho_\mu^\nu A_\nu -\frac {d+2}2\,  \Rho A_\mu\Big)=0\, ,
\ee
where 
$ A_\mu\equiv \wt V_\mu\lvert_{V^+=0}$ and transforms as $A_{\mu}\,\,\mapsto \Omega^{-\frac{d-4}2} A_\mu$. 
Having fused spin one theories in a single tractor equation, we now unify massive, partially massive and massless spin two systems.

\section{Spin Two}
We start with a weight $w$, rank two, symmetric tractor tensor $V^{MN}$ and propose its off-shell equations of motion and their gauge invariances:
 \bea 
I_R \Gamma^{RMN}\equiv G^{MN}=0\, ,\quad \nn \\[2mm] 
\delta V^{MN}=D^{(M}\xi^{N)}\, ,\qquad I\cdot \xi = 0\, ,\label{2gauge}
\eea
where the ``tractor Christoffel symbols'' are
 \be
2 \Gamma^{RMN}=2D^{(M} V^{N)R}- D^R V^{MN}\, .
 \ee
 This gauge invariance relies on a vanishing commutator of tractor $D$-operators \cite{GP} $[D^{M},D^{N}]=0$
 which necessitates conformally flat backgrounds.
 Since $V^{MN}$ has too high  a field content, we impose the gauge invariant constraint
\be
D\cdot V^N -\frac 12 D^N V^M_M=0\, .\label{Vc}
\ee
Picking a Weyl frame where the scale is constant and the metric is constant curvature, the off-shell equations yield three independent ones: $G^{++}$, $G^{+n}$, and $G^{mn}$.  Again any higher derivatives can be eliminated by taking certain linear combinations  $({\cal G}^{++},{\cal G}^{+m},{\cal G}^{mn})$.  The independent St\"uckelberg fields $V^{+m}$ and $V^{++}$ can be gauged away (so long as $w\neq0,-1$) giving a Pauli--Fierz equation in terms of a minimal covariant field content~$V_{\mu\nu}$:
\be
{\cal G}^{mn}\lvert_{V^{+}=0, V^{++}=0} \ = G_{\rm E}^{mn} + G_{\rm mass}^{mn}=0.\label{123}
\ee
The linearized cosmological Einstein tensor and Pauli--Fierz mass term~\cite{Pauli}~are 
\bea
G_{\rm E}^{mn} &=&\Big[ \Delta- \frac{4\Rho}{d} \Big]V^{mn}- 2\nabla^{(m} \nabla . V^{n)} \nn\\ &+&  \eta^{mn}\nabla.\nabla.V+ \nabla^m\nabla^n V^{r}_{r}
 -\eta^{mn}
\Big[\Delta+\frac{2\Rho(d-3)}{d}\Big]V^{r}_{r} \, ,\nn\\[2mm]
G_{\rm mass}^{mn} &=&  -m^2[V^{mn}- \eta^{mn}V^{r}_{r}]\, .
\eea
The mass-Weyl weight relationship and Breitenlohner--Freedman bound~are
\be
m^2=\frac{2\Rho}{d}\Big[\Big(\frac{d-1}{2}\Big)^2-\Big(w+\frac{d-1}{2}\Big)^2   \Big]\ \geq\  \frac{2\Rho}{d}\Big(\frac{d-1}{2}\Big)^2\, .\label{spin2mass}
\ee

Again we can analyze special Weyl weights:  At $ \w=0$, the Pauli-Fierz mass term~\eqn{spin2mass} vanishes to yield a theory of massless gravitons invariant under linearized diffeomorphisms.  At  $w=-1$, the Pauli-Fierz mass term is non-zero but a new double
derivative gauge invariance 
\be
\delta V^{mn}=\Big[\nabla^{m}\nabla^{n}+\frac{2\Rho}{d}\eta^{{mn}}\Big]\xi^{+}\, ,
\ee
appears and we recover the partially massless spin two theory~\cite{Deser:1983tm}.  At $w=1-d/2$ both of the field equations, $G^{++}$ and $G^{+m}$,  are zero and the scale decouples from $G^{MN}$.  This decoupling is the by now familiar phenomenon that we observed in scalar and spin one theories and produces a new conformally invariant but gauge variant theory with equation of motion
\be
\Gamma^{RMN}=0\, .
\ee
We now pass to analyzing systems with arbitrary higher spin values.

\section{Higher Spin Tractor Equations}
To describe higher spins in a unified way, we use a weight~$w$
(assumed generic), totally symmetric, rank~$s$, tractor $V^{M_1\cdots
  M_s}$ with the right number of constraints.  For constraints, we
impose \be 2 D\cdot V^{M_2\ldots M_s}-(s-1)D^{(M_2}_{\phantom M}
V_R^{M_3\ldots M_s)R}= 0=V_{RS}^{RSM_5\ldots M_s}\, ,\label{fc} \ee
while for equations of motion, we propose \be G^{M_1\ldots M_s}=
I\cdot D\, V^{M_1\ldots M_s} - s D^{(M_1}I\cdot V^{M_2\ldots M_s)}=0\,
, \label{arbit_spin} \ee with gauge invariances given by \be \delta
V^{M_1\ldots M_s}=D^{(M_1}\xi^{M_2\ldots M_s)}\, ,\label{sgauge}
\qquad I\cdot \xi^{M_1\ldots M_{s-2}} = 0 = \xi_R^{R M_3\ldots
  M_{s-1}}\, .  \ee This gauge invariance is predicated on conformally
flat backgrounds.  In~\cite{GoSW} we conjectured that a choice of gauge yields
the following on-shell equations: \be D\cdot V^{M_2\cdots M_s}= X
\cdot V^{M_2\cdots M_s}= I\cdot D\ V^{M_1\cdots M_s}= I\cdot
V^{M_2\cdots M_s} = V_R^{RM_3\cdots M_s}=0\, .
 \label {onshell}\ee
In a Weyl frame where $\sigma$ is constant these on-shell tractor equations are equivalent to the familiar on-shell equations for  massive spin~$s$ excitations propagating in constant curvature backgrounds (see the review~\cite{Bekaert:2005vh})
\be
\Big(\Delta +\frac{2\Rho}{d}[w(w+d-1)-s]\Big)V_{\mu_1\cdots\mu_s}=
\nabla.V_{\mu_2\ldots \mu_s}=
V^\rho_{\rho\mu_3\ldots\mu_s}=0\, .\label{osh}
 \ee
These  give the mass-Weyl weight relation and Breitenlohner--Freedman bound for an arbitrary spin $s$ 
\be
m^2=\frac{2\Rho}d \Big[ \Big(\frac{d-5}{2}+s\Big)^2-\Big(w+\frac{d-1}{2}\Big)^2\Big]
\ \geq\ \frac{2\Rho}{d}\, \Big(\frac{d-5}{2}+s\Big)^2\, .
\ee
Here we have chosen the zero of the mass parameter to coincide with the appearance of a
single derivative gauge invariance. 
Residual gauge invariances of these on-shell equations are discussed below.
A proof that they are equivalent to their tractor counterparts~(\ref{fc},\ref{arbit_spin}) at arbitrary~$s$ is reserved for 
a future publication~\cite{GoSW1} and relies on a tractor generalization of the symmetric form algebra
introduced and systemized in~\cite{Damour:1987vm,Hallowell:2005np,Labastida:1987kw,Vasiliev:1988xc,Duval,Duval1}. 
The $s=2$ proof is fairly simple: We pick the gauge (reachable at generic~$w$) where 
\be
X \cdot V^{N}=\sigma^{{-1}}\, X^{N} \,I\cdot X\cdot V \label{Tgauge}.
\ee
If we contract the field constraint~\eqn{fc} with $X$ and the equation of motion~\eqn{arbit_spin} twice with~$X$ and use
the gauge choice, equation~\eqn{identity} and the identity
\be
(d+2w-2)D_{MN} \equiv 2X_{[N}D_{M]} \,  ,\label{DD} 
\ee
we learn $V_{M}^{M}=I\cdot X\cdot V=0$ so that $X\cdot V^{N}=0$.
Then~\eqn{fc} implies $D\cdot V^{N}=0$.
To prove  $I\cdot D\ V^{MN}=0$, we contract~\eqn{arbit_spin} with $X$ once and use~(\ref{identity}), which implies         
$I\cdot V^{N} = 0$. This  relation along with~\eqn{arbit_spin} gives the desired result. 

Our final computation examines the partially massless gauge symmetries  enjoyed by constant
curvature higher spin systems for special tunings of their masses~\cite{Deser:2001pe}. These appear   as residual gauge symmetries of the on-shell equations~\eqn{onshell}  at special weights. For spins one and two they occur at $w=-1$ and $w=0,-1$, respectively;
a trend that holds for higher spins which  have residual gauge invariances at $w=s-2,s-3,\ldots,0,-1$ given by
\be
\delta V^{M_1\ldots M_s}= D^{(M_1}\cdots D^{M_t}\xi^{M_{t+1}\ldots M_s)}\, .
\ee
Here $\xi^{M_1\ldots M_{s-t}}$ is a weight $s-1$ parameter and $t$ is called the ``depth''
of a partially massless gauge transformation and $\xi$ is subject to 
$$
X\cdot \xi^{M_1\ldots M_{s-t-1}}=
I\cdot \xi^{M_1\ldots M_{s-t-1}}=
D\cdot \xi^{M_1\ldots M_{s-t-1}}=
$$
\be
I\cdot D\ \xi^{M_1\ldots M_{s-t}}=0=
\xi_R^{RM_1\ldots M_{s-t-2}}\, .
\ee
At depth $t$ the partially massless field $V^{M_1\ldots M_s}$   has weight $s-t-1$
and mass 
\be
m^2=\frac{2\Rho}d  (t-1)(d-4+2s-t)\, .
\ee
These results reproduce those found in~\cite{Deser:2001pe} by rather different methods.

\section{Conclusions}

In this Letter, we showed that massless, partially massless, and
massive theories can be unified in a single tractor equation.  The key
to this unification was understanding the origins of masses as Weyl
weights of tractors. This relationship led naturally to
Breitenlohner--Freedman stability bounds.  As detailed examples, we
explicitly constructed tractor equations for spin one and two.  For
spin one, it was shown that the Maxwell, Proca, and conformally
invariant theory of Deser-Nepemechie all spring from the same source.
The spin two tractor equations produced massless and partially
massless excitations in addition to massive spin two theories as well
as a novel conformally invariant, yet gauge variant theory existing in
any dimension.  It is striking that our techniques not only explain
the origins of mass, but also provide the technology to construct new
physical theories.

\section{Acknowledgements}
A.R.G. was supported by Marsden Grant no.\ 06-UOA-029 and a membership
of the Institute for Advanced Study Princeton.  A.W. is indebted to
the University of Auckland for warm hospitality.

\end{document}